\documentclass[runningheads]{llncs}
\usepackage{graphicx}
\usepackage{times}
\usepackage{amssymb}
\usepackage{latexsym}
\usepackage{amsmath,amsfonts}
\usepackage{hhline}
\renewcommand{\baselinestretch}{0.98}

\def\bldl{\begin{small}\vspace{-2mm}\[\bf\begin{array}{lcl}}
\def\cldl{\begin{small}\vspace{-0.4cm}\[\bf\begin{array}{lcl}}
\def\eldl{\end{array}\]\rm\end{small}}

\def\AC{\mbox{\cal AC}}

\begin{document}
\title{Consistent query answers on numerical databases\\ under aggregate constraints}
\titlerunning{CQA under aggregate constraints}
\author{Sergio Flesca \and Filippo Furfaro \and Francesco Parisi}
\institute{
DEIS - Universit\`a della Calabria\\
Via Bucci - 87036 Rende (CS) ITALY\\
Fax: +39 0984 494713\\
\email{\{flesca, furfaro, parisi\}@si.deis.unical.it}}
\authorrunning{S. Flesca, F. Furfaro,  F. Parisi}
\maketitle

\begin{abstract}
%\footnotesize
The problem of extracting consistent information from relational
databases violating integrity constraints on numerical data
is addressed.
In particular, aggregate constraints defined as linear
inequalities on aggregate-sum queries on input data are considered.
The notion of repair as consistent set of updates at attribute-value
level is exploited, and the characterization of several
complexity issues related to repairing data and computing consistent
query answers is provided.
%
%A value-update based framework for repairing data violating these
%forms of constraints is defined, and the characterization of several
%complexity issues related to repairing data and computing consistent
%query answers is provided.
\end{abstract}

%\vspace*{-3mm}
\section{Introduction}
Research has deeply investigated several issues related to the
use of integrity constraints on relational databases.
In this context, a great deal of attention has been devoted to
the problem of extracting reliable information from databases
containing pieces of information inconsistent w.r.t. some
integrity constraints.
All previous works in this area deal with ``classical" forms of
constraint (such as keys, foreign keys, functional dependencies),
and propose different strategies for updating inconsistent data
reasonably, in order to make it consistent by means of minimal
changes.
Indeed these kinds of constraint often do not suffice to manage
data consistency, as they cannot be used to define algebraic
relations between stored values.
In fact, this issue frequently occurs in several scenarios, such
as scientific databases, statistical databases, and data
warehouses, where numerical values of tuples are derivable
by aggregating values stored in other tuples.

In this work we focus our attention on databases where stored data
violates a set of \emph{aggregate constraints}, i.e. integrity
constraints defined on aggregate values extracted from the database.
These constraints are defined on numerical attributes (such as
sales prices, costs, etc.) which represent measure values and are
not intrinsically involved in other forms of constraints.

\begin{example}
Table \ref{tab:cashbudget} represents a two-years
\emph{cash budget} for a firm, that is a summary of cash flows
(receipts, disbursements, and cash balances) over the specified
periods.
Values `\emph{det}', `\emph{aggr}' and `\emph{drv}' in column \emph{Type}
stand for \emph{detail}, \emph{aggregate} and \emph{derived}, respectively.
In particular, an item of the table is \emph{aggregate} if it is obtained
by aggregating items of type \emph{detail} of the same section, whereas
a \emph{derived} item is an item whose value can be computed using the values of
other items of any type and belonging to any section.
%In this case, each Main Section has at most one aggregate item.
% DIRE CHE LA TABELLA PUO` ESSERE FACILMENTE OTTENUTA UTILIZZANDO
% STRUMENTI DI ACQUISIZIONE (SEMI-) AUTOMATICI. CITARE TOOL O ARTICOLI
% SULLA RICLASSIFICAZIONE DI BILANCIO

\begin{table}[!h]
\centering
\begin{tabular}{||l|l|l|c|r||}
  \hhline{|t:=====:t|}
  \raisebox{0cm}[3.2mm][1.1mm]{\textbf{\textit{Year}}} & \textbf{\textit{Section}}      & \textbf{\textit{Subsection}} & \textbf{\textit{Type}} & \textbf{\textit{Value}}\\
  \hhline{|:=====:|}
  \raisebox{0cm}[3mm][1mm]{2003} & Receipts       & beginning cash & drv &  20 \\
  \hhline{||-----||}
  \raisebox{0cm}[3mm][1mm]{2003} & Receipts       & cash sales     & det & 100\\
  \hhline{||-----||}
  \raisebox{0cm}[3mm][1mm]{2003} & Receipts       & receivables    & det & 120\\
  \hhline{||-----||}
  \raisebox{0cm}[3mm][1mm]{2003} & Receipts        &total cash receipts & aggr & 250\\
  \hhline{||-----||}
  \raisebox{0cm}[3mm][1mm]{2003} & Disbursements  &payment of accounts       & det & 120\\
  \hhline{||-----||}
  \raisebox{0cm}[3mm][1mm]{2003} & Disbursements  &capital expenditure & det &   0\\
  \hhline{||-----||}
  \raisebox{0cm}[3mm][1mm]{2003} & Disbursements  & long-term financing& det &  40\\
  \hhline{||-----||}
  \raisebox{0cm}[3mm][1mm]{2003} & Disbursements  &total disbursements & aggr & 160\\
  \hhline{||-----||}
  \raisebox{0cm}[3mm][1mm]{2003} & Balance         &net cash inflow     & drv &  60\\
  \hhline{||-----||}
  \raisebox{0cm}[3mm][1mm]{2003} & Balance         &ending cash balance & drv &  80\\
  \hhline{|:=====:|}
%  11& Cash Balance          &minimum cash                 & der &  10\\
%    &                       &balance                      &   &\\
%  \hline
%  12& Cash Balance          &surplus(deficit)             & der &  50\\
%  \hline
  \raisebox{0cm}[3mm][1mm]{2004} & Receipts       & beginning cash & drv &  80 \\
  \hhline{||-----||}
  \raisebox{0cm}[3mm][1mm]{2004} & Receipts       & cash sales     & det & 100\\
  \hhline{||-----||}
  \raisebox{0cm}[3mm][1mm]{2004} & Receipts       & receivables    & det & 100\\
  \hhline{||-----||}
  \raisebox{0cm}[3mm][1mm]{2004} & Receipts        &total cash receipts & aggr & 200\\
  \hhline{||-----||}
  \raisebox{0cm}[3mm][1mm]{2004} & Disbursements  &payment of accounts & det & 130\\
  \hhline{||-----||}
  \raisebox{0cm}[3mm][1mm]{2004} & Disbursements  &capital expenditure & det &  40\\
  \hhline{||-----||}
  \raisebox{0cm}[3mm][1mm]{2004} & Disbursements  & long-term financing & det &  20\\
  \hhline{||-----||}
  \raisebox{0cm}[3mm][1mm]{2004} & Disbursements  &total disbursements & aggr & 190\\
  \hhline{||-----||}
  \raisebox{0cm}[3mm][1mm]{2004} & Balance         &net cash inflow     & drv &  10\\
  \hhline{||-----||}
  \raisebox{0cm}[3mm][1mm]{2004} & Balance         &ending cash balance & drv &  90\\
\hhline{|b:=====:b|}
\end{tabular}
\label{tab:cashbudget}
\vspace*{0,3 cm}\caption{A cash budget}
\end{table}

A cash budget must satisfy these integrity constraints:
\begin{list}{-}{\leftmargin .1cm \itemsep 0cm \parsep 0cm}
\item[1.]
for each section and year, the sum of the values of all \emph{detail}
items must be equal to the value of the \emph{aggregate} item of the same
section and year;
\item[2.]
for each year, the net cash inflow must be equal to the difference
between total cash receipts and total disbursements;
\item[3.]
for each year, the ending cash balance must be equal to the sum of the
beginning cash and the net cash balance.
\end{list}

Table \ref{tab:cashbudget} was acquired by means of an OCR
tool from two paper documents, reporting the cash budget
for $2003$ and $2004$.
The original paper document was consistent, but some symbol recognition
errors occurred during the digitizing phase, as constraints 1) and 2)
are not satisfied on the acquired data for year $2003$, that is:
\begin{list}{-}{\leftmargin .2cm \itemsep 0cm \parsep 0cm}
\item[i)]
in section \emph{Receipts}, the aggregate value of
\emph{total cash receipts} is not equal to the sum of detail values
of the same section.%: $100+120 \neq 250$;
\item[ii)]
the value of \emph{net cash inflow} is not to equal the difference between
\emph{total cash receipts} and \emph{total disbursements}.%: $60 \neq 250-160$.
\end{list}

In order to exploit the digital version of the cash budget, a
fundamental issue is to define a reasonable strategy for locating
OCR errors, and then ``repairing" the acquired data to extract
reliable information.
\label{ex:esempiointro}
\end{example}

Most of well-known techniques for repairing data violating either key
constraints or functional dependencies accomplish this task by performing
deletions and insertions of tuples.
Indeed this approach is not suitable for contexts analogous to that
of Example~\ref{ex:esempiointro}, that is of data acquired by OCR
tools from paper documents.
For instance, repairing Table \ref{tab:cashbudget} by either
adding or removing rows means hypothesizing that the OCR tool either
jumped a row or ``invented" it when acquiring the source paper document,
which is rather unrealistic.
The same issue arises in other scenarios dealing with numerical data
representing pieces of information acquired automatically, such as
sensor networks.
In a sensor network with error-free communication channels, no reading
generated by sensors can be lost, thus repairing the database by adding
new readings (as well as removing collected ones) is of no sense.
In this kind of scenario, the most natural approach to data repairing is
updating directly the numerical data: this means working at attribute-level,
rather than at tuple-level.
For instance, in the case of Example \ref{ex:esempiointro}, we can reasonably
assume that inconsistencies of digitized data are due to symbol recognition
errors, and thus trying to re-construct actual data values is well founded.
Likewise, in the case of sensor readings violating aggregate constraints,
we can hypothesize that inconsistency is due to some trouble occurred
at a sensor while generating some reading, thus repairing data by modifying
readings instead of deleting (or inserting) them is justified.

%Inserimenti e cancellazioni riescono a riparare i dati inconsistenti
%rispetto a vincoli di chiave, ma nei contesti appena menzionati possiamo
%lecitamente supporre che non vi siano incongruenze rispetto a tali vincoli.
%Noi poniamo la nostra attenzione su dati numerici che sono essenzialmente
%valori di misura e che non sono intrinsecamente coinvolti in altri tipi
%di vincoli.

% OVVIAMENTE IN QUESTI CONTESTI E' LECITO SUPPORRE CHE NON CI SIANO
% INCONGRUENZE RISPETTO AD ALTRE CLASSI DI VINCOLI: OSSIA I DATI CHE
% CONSIDERIAMO SONO CONSISTENNTI RISPETTO A VINCOLI DI CHIAVE, CHIAVE
% ESTERNA, ECC.
%\vspace*{-1mm}
\subsection{Related Work}
%\vspace*{-1mm}
First theoretical approaches to the problem of dealing with incomplete
and inconsistent information date back to 80s, but these works mainly
focus on issues related to the semantics of incompleteness \cite{Imi*84}.
The problem of extracting reliable information from inconsistent data
was first addressed in \cite{Aga*95}, where an extension of relational
algebra (namely \emph{flexible algebra}) was proposed to evaluate queries
on data inconsistent w.r.t. key constraints (i.e. tuples having the same
values for key attributes, but conflicting values for other attributes).
% Si fondono le tuple contrastanti in un'unica tupla definita su uno
% schema non 1NF, ossia la fusione crea attributi non ``piatti", ma
% aventi insiemi per valori
% In \cite{Dung97} si mostra che la tecnica di \cite{Aga*95} non funziona su
% chiavi multiple (ossia se cisono più chiavi candidate) and an extension of
% flexible algebra (namely \emph{Integrated Relational Calculus}) was introduced.
The first proof-theoretic notion of \emph{consistent query answer} was
introduced in \cite{Bry*97}, expressing the idea that tuples involved
in an integrity violation should not be considered in the evaluation of
consistent query answering.
In \cite{Are*99} a different notion of consistent answer was introduced,
based on the notion of \emph{repair}:
a repair of an inconsistent database $D$ is a database $D'$ satisfying
the given integrity constraints and which is minimally different from $D$.
Thus, the consistent answer of a query $q$ posed on $D$ is the answer
which is in every result of $q$ on each repair $D'$.
% Si fondono le tuple contrastanti in un'unica tupla definita su uno
% schema non 1NF, ossia la fusione crea attributi non ``piatti", ma
% aventi insiemi per valori
% In \cite{Dung97} si mostra che la tecnica di \cite{Aga*95} non funziona
% su chiavi multiple (ossia se cisono più chiavi candidate) and an
% extension of flexible algebra (namely \emph{Integrated Relational Calculus})
% was introduced.
In particular, in \cite{Are*99} the authors show that, for restricted classes
of queries and constraints, consistent answers can be evaluated without
computing repairs, but by looking only at the specified constraints and
rewriting the original query $q$ into a query $q'$ such that the answer
of $q'$ on $D$ is equal to the consistent answer of $q$ on $D$.
Based on the notions of repair and consistent query answer introduced
in \cite{Are*99}, several works investigated more expressive classes of
queries and constraints.
In \cite{Are*00} extended disjunctive logic programs with exceptions were
used for the computation of repairs, and in \cite{Are*03} the evaluation of
aggregate queries on inconsistent data was investigated.
A further generalization was proposed in \cite{Gre*03}, where the authors
defined a technique based on the rewriting of constraints into extended
disjunctive rules with two different forms of negation (negation as failure
and classical negation).
This technique was shown to be sound and complete for universally quantified
constraints.

All the above-cited approaches assume that tuple insertions
and deletions are the basic primitives for repairing inconsistent data.
More recently, in \cite{Cho*05} a repairing strategy using only tuple
deletions was proposed, and in \cite{Wij*03} repairs also consisting of
update operations were considered.
The latter is the first approach performing repairs at the attribute-value
level, but is not well-suited in our context, as it works only in the case
that constraints consist of full dependencies.

The first work investigating aggregate constraints on numerical data is
\cite{Ros*98}, where the consistency problem of very general forms of
aggregation is considered, but no issue related to data-repairing is
investigated.
In \cite{Bert*05} the problem of repairing databases by fixing numerical
data at attribute level is investigated.
The authors show that deciding the existence of a repair under both
denial constraints (where built-in comparison predicates are allowed)
and a non-linear form of multi-attribute aggregate constraints is
undecidable.
Then they disregard aggregate constraints and focus on the problem of
repairing data violating denial constraints, where no form of aggregation
is allowed in the adopted constraints.

% This is mainly due to the fact that the presence of multi-attribute
% allow arithmetical combinations of attributes in the aggregation function.
% The other results in the latter work involve only denial constraints in the
% context of numerical data.

% si potrebbe aggiungere che la misura usata tra D e D' è la differenza
% tra i quadrati dei valori presenti prima e dopo il repair ;
% L'adozione di questa metrica non è giustibicabile ....

%To the best of our knowledge there is no work in literature which considers
%integrity violations w.r.t. ``numerical" constraints.

\vspace*{-1mm}
\subsection{Main Contribution}
\vspace*{-1mm}
We investigate the problem of repairing and extracting reliable
information from data violating a given set of aggregate constraints.
These constraints consist of linear inequalities on
aggregate-sum queries issued on measure values stored in the database.
This syntactic form enables meaningful constraints to be expressed,
such as those of Example \ref{ex:esempiointro} as well as other forms
which often occur in practice.

We consider database repairs consisting of ``reasonable" sets of
value-update operations aiming at re-constructing the correct
measure values of inconsistent data.
We adopt two different criteria for determining whether a set of update
operations repairing data can be considered ``reasonable" or not:
\emph{set}-minimal semantics and \emph{card}-minimal semantics.
Both these semantics aim at preserving the information represented in
the source data as much as possible.
They correspond to different repairing strategies which turn out to be
well-suited for different application scenarios.

We provide the complexity characterization of three fundamental
problems:
i) \emph{repairability} (is there at least one repair for the given
database w.r.t. the specified constraints?);
ii) \emph{repair checking} (given a set of update operations, is it a
``reasonable" repair?);
iii) \emph{consistent query answer} (is a given boolean query true in
every ``reasonable" repair?).

\section{Preliminaries}\label{sec:prel}
%\vspace*{-1mm}
We assume classical notions of database scheme, relational scheme,
and relations.
In the following we will also use a logical formalism to represent
relational databases, and relational schemes will be represented
by means of sorted predicates of the form
$R(A_1\!:\!\Delta_1, \dots, A_n\!:\!\Delta_n)$, where $A_1, \dots, A_n$
are attribute names and $\Delta_1, \dots, \Delta_n$ are the corresponding
domains.
Each $\Delta_i$ can be either $\mathbb{Z}$ (infinite domain of integers),
$\mathbb{R}$ (reals), or $\mathbb{S}$ (strings).
% and will be denoted as \emph{relational domains}.
Domains $\mathbb{R}$ and $\mathbb{Z}$ will be said to be
\emph{numerical domains}, and attributes defined over $\mathbb{R}$ or
$\mathbb{Z}$ will be said to be \emph{numerical attributes}.
Given a ground atom $t$ denoting a tuple, the value of attribute
$A$ of $t$ will be denoted as $t[A]$.

Given a database scheme $\mathcal{D}$, we will denote as $\mathcal{M_D}$
(namely, \emph{Measure attributes}) the set of numerical attributes
representing measure data.
%
%which are not ``involved" in neither key nor foreign key constraints.
%That is,
%%if $D$ is updated by changing the values of only
%%attributes in $S_D$, then $D$ will be still consistent w.r.t.
%%these constraints.
%any update $u$ involving only attributes in $S_D$ is ``safe", in the
%sense that $D$ is still consistent w.r.t. the specified keys and foreign
%keys after $u$ is performed.
That is, $\mathcal{M_D}$ specifies the set of attributes representing measure
values, such as weights, lengths, prices, etc.
For instance, in Example \ref{ex:esempiointro}, $\mathcal{M_D}$ consists of
the only attribute \emph{Value}.

Given two sets $M$, $M'$, $M \triangle M'$  denotes their symmetric
difference $( M \cup M' ) \setminus ( M \cap M' )$.

\subsection{Aggregate constraints}

%Given a relational scheme $R$ in the database $D$, let
%$\mathcal{M}_R=\{A_1, \dots, A_k \}$ be the subset of $\mathcal{M_D}$ containing
%all the attributes in $R$ belonging to $\mathcal{M_D}$.
Given a relational scheme $R(A_1\!:\!\Delta_1, \dots, A_n\!:\!\Delta_n)$,
an \emph{attribute expression} on $R$ is defined recursively as follows:
%\vspace*{-1mm}
\begin{list}{-}{\leftmargin .2cm \itemsep 0mm}
\item
a numerical constant is an attribute expression;
\item
each $A_i$ (with $i\in [1..n]$) is an attribute expression;
\item
$e_1 \psi e_2$ is an attribute expression on $R$, if $e_1$, $e_2$ are
attribute expressions on $R$ and $\psi$ is an arithmetic operator in
$\{+, - \}$;
\item
$c\!\times\!(e)$ is an attribute expressions on $R$,
if $e$ is an attribute expression on $R$ and $c$ a
numerical constant.
%\item
%$(e)$ and $-(e)$ are attribute expressions on $R$, if $e$ is an attribute expression on $R$.
\end{list}

%The numerical value obtained evaluating an expression $e$ on a tuple $t$ will be denoted
%as $e(t)$.
%Given a relation $r$ on $R$ and an attribute expression $e$ on $R$,
%we denote as $e(t)$ the numerical value obtained evaluating $e$ on the tuple $t$ in $r$.

Let $R$ be a relational scheme, $e$ an attribute expression on $R$,
and $C$ a boolean formula on constants and attributes of $R$.
An \emph{aggregation function} on $R$ is a function
$\chi:(\Delta_{1}\times \dots\times\Delta_{k})\rightarrow \mathbb{R}$,
where $\Delta_{1}, \dots, \Delta_{k}$ are the relational domains of some
attributes $A_{1}, \dots, A_{k}$ of $R$.
$\chi(x_1,\dots ,x_k)$ is defined as follows:

\noindent%\hspace*{-2mm}%\vspace*{-2mm}
{\small
$
\begin{array}{lll}
   \mbox{\normalsize $\chi(x_1,\dots ,x_k)$} = & \tt SELECT & \tt sum(e) \\
                         & \tt FROM   & \tt R \\
                         & \tt WHERE  & \tt \alpha(x_1,\dots,x_k)
\end{array}
$}\\
where $\alpha(x_1,\dots,x_k) = C \wedge\ (\,A_{1}\!=\!x_1\,)\ \wedge \dots \wedge\ (\, A_{k}\!=\!x_k\,)$.

\noindent

\begin{example}
The following aggregation functions are defined on the relational scheme
\emph{CashBudget(Year, Section, Subsection, Type, Value)} of
Example \ref{ex:esempiointro}:\vspace*{1mm}

\noindent
{\small
$
\begin{array}{cc}
\begin{array}{lll}
 \mbox{\normalsize $\chi_1(x,y,z)$} = & \tt SELECT & \tt sum(Value) \\
                   & \tt FROM   & \tt CashBudget \\
                   & \tt WHERE  &\tt Section\!=\!x \ \\
                   & \ \ \ \ \ \ \tt AND &\tt Year\!=\!y \ \tt AND \ \tt Type\!=\!z
\end{array}
&
\hspace*{4mm}
\begin{array}{lll}
  \mbox{\normalsize $\chi_2(x,y)$} = & \tt SELECT & \tt sum(Value) \\
                   & \tt FROM   & \tt CashBudget \\
                   & \tt WHERE  & \tt  Year=x  \\
                   & \ \ \ \ \ \ \tt AND & \tt Subsection\!=\!y
\end{array}
\end{array}
$}

%\vspace*{1mm}
\vspace*{1mm}\noindent
Function $\chi_1$ returns the sum of \emph{Value} of all the tuples having
\emph{Section} $x$, \emph{Year} $y$ and \emph{Type} $z$.
For instance, $\chi_1(\mbox{\small `Receipts', `2003', `det'})$ returns
$100+120=$ $220$, whereas $\chi_1(\mbox{\small `Disbursements', `2003', `aggr'})$
returns $160$.
Function $\chi_2$ returns the sum of \emph{Value} of all the tuples
where \emph{Year=x} and \emph{Subsection=y}.
In our running example, as the pair \emph{Year}, \emph{Subsection} uniquely
identifies tuples of \emph{CashBudget}, the sum returned by $\chi_2$ coincides
with a single value.
For instance, $\chi_2(\mbox{`2003', `cash sales'})$ returns $100$,
whereas $\chi_2(\mbox{`2004', `net cash inflow'})$ returns $10$.
\label{ex:esempiofunzioni}
\end{example}

%\begin{example}
%The following aggregation function is defined on the relational scheme
%\emph{CashBudget(Year, Section, Subsection, Type, Value)} of
%Example \ref{ex:esempiointro} and returns the sum of all items of
%\emph{Section} $x$, \emph{Year} $y$ and \emph{Type} $z$:
%
%%\vspace*{1mm}\hspace*{-3mm}
%\noindent
%{\small
%$
%\begin{array}{lll}
% \mbox{\normalsize $\chi_1(x,y,z)$} = & \tt SELECT & \tt sum(Value) \\
%                   & \tt FROM   & \tt CashBudget \\
%                   & \tt WHERE  &\tt Section\!=\!x \ \wedge \ Year\!=\!y \wedge \ Type\!=\!z
%\end{array}
%$}
%
%%\vspace*{1mm}
%\noindent
%For instance, $\chi_1(\mbox{\small `Receipts', `2003', `det'})$ returns $100+120=$ $220$,
%whereas $\chi_1(\mbox{\small `Disbursements', `2003', `aggr'})$ returns $160$.
%
%The following aggregation function returns attribute  \emph{Value} of the item where \emph{Year=x} and
%\emph{Subsection=y}:
%
%%\vspace*{1mm}
%\noindent
%{\small
%$
%\begin{array}{lll}
%  \mbox{\normalsize $\chi_2(x,y)$} = & \tt SELECT & \tt sum(Value) \\
%                   & \tt FROM   & \tt CashBudget \\
%                   & \tt WHERE  & \tt  Year=x  \wedge \ Subsection=y;
%\end{array}
%$
%}
%
%%\vspace*{1mm}
%\noindent
%For instance, $\chi_2(\mbox{`2003', `cash sales'})$ returns $100$,
%whereas $\chi_2(\mbox{`2004', `net cash inflow'})$ returns $10$.
%\label{ex:esempiofunzioni}
%\end{example}

\begin{definition}[Aggregate constraint]
Given a database scheme $\mathcal{D}$, an aggregate
constraint on $\mathcal{D}$ is an expression of the form:
%\vspace*{-3mm}
{%\small
\begin{equation}\label{eq:aggrConstraint}
%\begin{center}
\forall x_1, \dots, x_k  \ \left(\phi(x_1,\dots,x_k) \implies \sum_{i=1}^n c_i\cdot \chi_i(X_i) \leq K\right)
%\end{center}
\end{equation}}

\vspace*{-3mm}
\noindent
where:
\vspace*{-1mm}
\begin{list}{-}{\leftmargin .2cm \itemsep -0mm}
\item[1.]
$c_1, \dots, c_n, K$ are constants;
\item[2.]
$\phi(x_1,\dots,x_k)$ is a conjunction of atoms containing the variables
$x_1,\dots,x_k$;
\item[3.]
each $\chi_i(X_i)$ is an aggregation function, where
$X_i$ is a list of variables and constants, and variables appearing in
$X_i$ are a subset of $\{x_1, \dots, x_k\}$.
\end{list}
\end{definition}

%A \emph{set of aggregate constraints} is denoted as $\mathcal{AC}$.
Given a database $D$ and a set of aggregate constraints
$\mathcal{AC}$, we will use the notation $D \models \mathcal{AC}$
[resp. $D \not\models \mathcal{AC}$] to say that $D$ is consistent
[resp. inconsistent] w.r.t. $\mathcal{AC}$.\\
Observe that aggregate constraints enable equalities to be expressed as well,
since an equality can be viewed as a pair of inequalities.
For the sake of brevity, in the following equalities will be written explicitly.

\begin{example}
%Consider the database scheme $\mathcal{D}=\{ \mbox{\emph{CashBudget}} \}$,
%where \emph{CashBudget} is the relational scheme of Example
%\ref{ex:esempiofunzioni}.
Constraint $1$ defined in Example \ref{ex:esempiointro} can be expressed
as follows:\\
$
\begin{array}{c}
 \forall \ x, y, s, t, v  \quad CashBudget( y, x, s, t, v) \implies  \chi_1(x,y,\mbox{`det'})- \chi_1(x,y,\mbox{`aggr'})= 0
\end{array}
$
\label{ex:vincoliAggregatiNotazione}
\end{example}
For the sake of simplicity, in the following we will use a shorter notation
for denoting aggregate constraints, where universal quantification is implied
and variables in $\phi$ which do not occur in any aggregation function are
replaced with the symbol `\_'.
For instance, the constraint of Example \ref{ex:vincoliAggregatiNotazione}
can be written as follows:\\
$
\begin{array}{c}
CashBudget( y, x, \_\ , \_\ , \_) \implies \chi_1(x,y,\mbox{`\emph{det}'})- \chi_1(x,y,\mbox{`\emph{aggr}'})  = 0
\end{array}
$

\begin{example}
\label{ex:vincoliAggregati}
Constraints 2 and 3 defined in Example \ref{ex:esempiointro} can be
expressed as follows:\\
\noindent
{\small
$
\begin{array}{l}
\mbox{\normalsize Constraint 2:} \hspace*{5mm}CashBudget ( x, \_\ , \_\ , \_\ , \_ )   \implies  \\
\hspace*{5mm}
\chi_2(x, \mbox{`net cash inflow'}) -
\left(
\chi_2(x, \mbox{`total cash receipts'}) -
\chi_2(x, \mbox{`total disbursements'})
\right)=0
\end{array}
$
}

\noindent
{\small
$
\begin{array}{l}
\mbox{\normalsize Constraint 3:} \hspace*{5mm}CashBudget ( x, \_\ , \_\ , \_\ , \_ )   \implies  \\
\hspace*{5mm}
\chi_2(x, \mbox{`ending cash balance'})-
\left(
\chi_2(x, \mbox{`beginning cash'})+
\chi_2(x, \mbox{`net cash balance'})
\right)=0
\end{array}
$
}

\vspace*{1mm}
\noindent
Consider the database scheme consisting of relation \emph{CashBudget}
and relation \emph{Sales( Product, Year, Income)}, containing pieces
of information on annual product sales.
The following aggregate constraint says that, for each year,
the value of \emph{cash sales} in \emph{CashBudget} must be equal
to the total incomes obtained from relation \emph{Sales}:\\
\vspace{1mm}{
{%\small
$
\begin{array}{c}
\mbox{CashBudget} \quad  ( x, \_\ , \_\ , \_\ , \_) \wedge \ \mbox{Sales}(\_\ , x, \_)
\implies \chi_2(x,\mbox{`cash sales'})- \chi_3(x)  = 0
\end{array}
$
}
}\\
\noindent
where $\chi_3(x)$ is the aggregation function returning the total income
due to products sales in year $x$:\\
%\vspace*{1mm}
\noindent
{\small $
\begin{array}{lll}
   \mbox{\normalsize $\chi_3(x)$} = & \tt SELECT & \tt sum(Income) \\
            & \tt FROM   & \tt Sales \\
            & \tt WHERE  & \tt Year =x \\
\end{array}
$}
\end{example}

%Our constraints should be considered orthogonal to key and
%foreign key constraints, as aggregation functions compute aggregate
%values of measure attributes, which are not intrinsically involved in
%neither key nor foreign key constraints.

%For instance, in our running example, neither does attribute
%\emph{Value} identify tuples of \emph{CashBudget}, nor might it be
%forced to take values from another key attribute.

%Therefore in the following we assume that input data is consistent
%w.r.t. key and foreign key constraints (otherwise, input data can be
%made consistent by means of one of well-known techniques);
%we focus our attention on the problem of updating numerical values
%to ...

\subsection{Updates}
Updates at attribute-level will be used in the following as the
basic primitives for repairing data violating aggregate constraints.
Given a relational scheme $R$ in the database scheme $\mathcal{D}$, let
$\mathcal{M}_R=\{A_1, \dots, A_k \}$ be the subset of $\mathcal{M_D}$ containing
all the attributes in $R$ belonging to $\mathcal{M_D}$.

\begin{definition}[Atomic update]
Let $t=R(v_1, \dots, v_n)$ be a tuple on the relational scheme
$R(A_1\!:\!\Delta_1, \dots, A_n\!:\!\Delta_n)$.
An \emph{atomic update} on $t$ is a triplet $<t, A_i, v'_i>$, where
$A_i \in \mathcal{M}_R$ and $v'_i$ is a value in $\Delta_i$ and $v'_i\neq v_i$.
\end{definition}

Update $u= <t, A_i, v'_i>$ replaces $t[A_i]$ with $v'_i$, thus yielding
the tuple $u(t)=R(v_1, \dots, v_{i-1}, v'_i, v_{i+1}, \dots, v_n)$.

Observe that atomic updates work on the set $\mathcal{M}_R$ of measure
attributes, as our framework is based on the assumption that data inconsistency
is due to errors in the acquisition phase (as in the case of digitization
of paper documents) or in the measurement phase (as in the case of
sensor readings).
Therefore our approach will only consider repairs aiming at re-constructing
the correct measures.
%
%
%
%measurement troubles or
%safely updatable
%attributes of the given relational scheme.
%That is, we assume that updates can be performed only on attributes
%which do not define neither key nor foreign key constraints.
%In fact, this is a well-founded assumption, as aggregation functions
%invoked in our form of constraint compute aggregate values of measure
%attributes, which are not intrinsically involved in any other forms of
%constraints.
%
%
%
%
%
%Therefore our form of constraints should be considered as orthogonal to
%key and foreign key constraints,
%
%and thus in the following we will not
%care of inconsistencies w.r.t. keys and foreign keys
%Therefore our form of constraints is orthogonal to key and foreign key
%constraints, so that updating values of attributes aggregated in our constraints
%cannot yield violations of either keys and foreign keys.
%
%
%Therefore, in the following we will not consider the presence of keys and foreign keys,
%
%
%our form of constraints is orthogonal to
%key and foreign key constraints.
%In fact, this is well-founded in real-life scenarios, as aggregation
%in practice is performed on attributes INTRINSECAMENTE DI MISURA.
%Therefore, updating values of attributes aggregated in our constraints
%cannot yield violations of either keys and foreign keys.

\begin{example}
\label{ex:esempioupdate}
Update $u=<t, \mbox{\emph{Value}}, 130 >$ issued on tuple
$t=\mbox{CashBudget}(2003,$ $\mbox{Receipts}, \mbox{cash sales}, \mbox{det}, 100)$
returns $u(t)=\mbox{CashBudget}(\mbox{2003, Receipts, cash sales,}$  $\mbox{det, 130})$.
%Consider the tuple $t=\mbox{CashBudget}(2003,$ $\mbox{Receipts}, \mbox{cash sales}, \mbox{det}, 100)$
%in the relation depicted in Fig. \ref{fig:cashbudget}.
%Update $u=<t, \mbox{\emph{Value}}, 130 >$ returns .
\end{example}

Given an update $u$, we denote the attribute updated by $u$ as $\lambda(u)$.
That is, if $u =$ $ <~t,~ A_i,~ v>$ then $\lambda(u) = <t, A_i>$.

\begin{definition}[Consistent database update]
Let $D$ be a database and $U=\{u_1, \dots, u_n\}$ be a set of atomic updates
on tuples of $D$. %, where $u_j=<t_j, A_j, v'_{i_j}>$ for each $j \in [1..n]$.
The set $U$ is said to be a \emph{consistent database update} iff
$\ \forall \ j, k \in [1..n]$ if $j\!\neq\!k$ then $\lambda(u_j)\neq \lambda(u_k)$.
\end{definition}

Informally, a set of atomic updates $U$ is a consistent database update
iff for each pair of updates $u_1, u_2 \in U$, $u_1$ and $u_2$ do not
work on the same tuples, or they change different attributes of the same
tuple.

The set of pairs $<\mbox{\emph{tuple}}, \mbox{\emph{attribute}}>$ updated
by a consistent database update $U$ will be denoted as
$\lambda(U)=\cup_{u_i\in U} \lambda(u_i)$.
%, and the number of values replaced by $U$ as $\delta(U)=|\lambda(U)|$.

Given a database $D$ and a consistent database update $U$, the result of
performing $U$ on $D$ consists in the new database $U(D)$ obtained by
performing all atomic updates in $U$.

\section{Repairing inconsistent databases}
\begin{definition}[Repair]
Let $\mathcal{D}$ be a database scheme, $\mathcal{AC}$ a set of
aggregate constraints on $\mathcal{D}$, and $D$ an instance
of $\mathcal{D}$ such that $D\not\models\mathcal{AC}$.
A \emph{repair $\rho$} for $D$ is a consistent database update such
that $\rho(D)\models\mathcal{AC}$.
\end{definition}

\begin{example}
\label{ex:primoesempiorepair}
A repair $\rho$ for \emph{CashBudget} w.r.t. constraints 1), 2) and 3)
consists in decreasing attribute \emph{Value} in the tuple
$t= \mbox{CashBudget}(\mbox{2003, Receipts, total cash receipts,}$ $\mbox{aggr, 250})$
down to 220;
that is, $\rho=\{\ <t, \mbox{\emph{Value}}, 220> \ \}$.
\end{example}

We now characterize the complexity of the repair-existence problem.
All the complexity results in the paper refer to data-complexity,
that is the size of the constraints is assumed to be bounded by a constant.

The following lemma is a preliminary result which states that potential
repairs for an inconsistent database can be found among set of updates
whose size is polynomially bounded by the size of the original database.

\begin{lemma}
\label{lem:dimpolrepair}
Let $\mathcal{D}$ be a database scheme,
$\mathcal{AC}$ a set of aggregate constraints on $\mathcal{D}$,
and $D$ an instance of $\mathcal{D}$ such that $D\not\models\mathcal{AC}$.
If there is a repair $\rho$ for $D$ w.r.t. $\mathcal{AC}$, then there is
a repair $\rho'$ for $D$ such that $\lambda(\rho')\subseteq\lambda(\rho)$
and $\rho'$ has polynomial size w.r.t. $D$.
\end{lemma}
%\bproof
\begin{proof}
(sketch)
W.l.o.g. we assume that the attribute expression $e_{\chi_i}$ occurring
in each aggregate function $\chi_i$ in $\mathcal{AC}$ is either an
attribute or a constant.
Let $\rho$ be a repair for $D$, and $\mathcal{AC}^{*}$ be the set of
inequalities obtained as follows:
\begin{list}{-}{\leftmargin .3cm \itemsep 0cm \parsep 0cm}
\item[1.]
a variable $x_{t, A}$ is associated to each pair
$<t, A>\ \in \lambda(\rho)$;
\item[2.]
for every constraint in $\mathcal{AC}$ of the form (\ref{eq:aggrConstraint})
and for every ground substitution $\theta$ of $x_1, \dots, x_k$
s.t. $\phi(\theta x_1, \dots, \theta x_k)$ is true, the following
inequalities are added to $\mathcal{AC}^{*}$:
\begin{list}{-}{\leftmargin .2cm \itemsep 0cm \parsep 0cm}
\item[a.]
$\sum_{i=1}^n c_i\cdot \sum_{<t, e_{\chi_i}>\ \in \lambda(\rho)\,\wedge\,t\models\,
\alpha_i(\theta x_1, \dots,\theta  x_k)} x_{t, e_{\chi_i}} \leq K'$,
where $K'$ is $K$ minus the contribution to the left-hand side of the
constraint due to values which have not been changed by $\rho$, i.e.
%we assume without loss of generality that each attribute expression
%defining the result of a aggregation function is either an attribute
%$A_i$ or a constant and
$K' = K - \sum_{i=1}^n c_i\cdot \sum_{<t,e_{\chi_i}>\ \not\in \lambda(\rho)\,\wedge\,t\models\,
\alpha_i(\theta x_1, \dots,\theta  x_k)} e_{\chi_i}$.
\item[b.]
for each tuple $t$ such that
$t\models\, \alpha_i(\theta x_1, \dots,\theta  x_k)$, let $\alpha'_i$
be the disjunctive normal form of $\alpha_i$ and let $\beta$ be a disjunct
in $\alpha'_i$ such that $t\models \beta(\theta x_1, \dots,\theta  x_k)$.
For each conjunct $\gamma$ in $\beta$ of the form $w_1\, \diamond\, w_2$,
where $\diamond$ is a comparison operator, and either $w_1$ or $w_2$ is
an attribute $A$ such that $<t,A>\in\lambda(\rho)$,
the constraint $v_1 \diamond v_2$ is added to  $\mathcal{AC}^*$,
where, for $\j \in \{1,2\}$ $^{1)}$ if $w_j$ is constant, $v_j = w_j$;
$^{2)}$ if $w_j = A$ and $<t, A>\in\lambda(\rho)$, $v_j = x_{t,A}$;
$^{3)}$ if $w_j = A$ and $<t, A>\not\in\lambda(\rho)$, $v_j = t[A]$.
\end{list}
\end{list}

Obviously $\mathcal{AC}^*$ has one solution, which corresponds to
assigning to each variable $x_{t,A_i}$ the value assigned by $\rho$
to attribute $A_i$ of tuple $t$.
Moreover, the number of variables and equations, and the size of constants
in $\mathcal{AC}^*$ are polynomially bounded by the size of $D$.
Therefore there is a solution $\overline{X}$ to $\mathcal{AC}^*$ whose size is
polynomially bounded by the size of $D$, since $\mathcal{AC}^*$ is a $PLI$
problem with at least one solution \cite{Pap81}.
$\overline{X}$ defines a repair $\rho'$ for $D$ such that
$\lambda(\rho')\subseteq\lambda(\rho)$ and $\rho'$ has polynomial size
w.r.t. $D$.
\qed
\end{proof}
%\eproof

\begin{theorem}[Repair existence]
\label{theo:repexistence}
Let $\mathcal{D}$ be a database scheme,
$\mathcal{AC}$ a set of aggregate constraints on $\mathcal{D}$,
and $D$ an instance of $\mathcal{D}$ such that $D\not\models\AC$.
The problem of deciding whether there is a repair for $D$ is NP-complete.
\end{theorem}
%\bproof
\begin{proof}
\emph{Membership.}
A polynomial size witness for deciding the existence of a repair is a
database update $U$ on $D$: testing whether $U$ is a repair for $D$
means verifying $U(D)\models \mathcal{AC}$, which can be accomplished
in polynomial time w.r.t. the size of $D$ and $U$.
If a repair exists for $D$, then Lemma \ref{lem:dimpolrepair}
guarantees that a polynomial size repair for $D$ exists too.\\
\emph{Hardness.}
We show a reduction from \textsc{circuit sat} to our problem.
Without loss of generality, we consider a boolean circuit $C$ using
only NOR gates.
The inputs of $C$ will be denoted as $x_1, \dots, x_n$.
The boolean circuit $C$ can be represented by means of the database
scheme:
\vspace*{-2mm}
\begin{quote}
$gate(\underline{IDGate},norVal,orVal),\,$\\
$gateInput(\underline{IDGate,IDIngoing},Val)$,\\
$input(\underline{IDInput},Val)$.\vspace*{-2mm}
\end{quote}
Therein:\vspace*{-2mm}
\begin{enumerate}
\item
each gate in $C$ corresponds to a tuple in $gate$
(attributes \emph{norVal} and \emph{orVal} represent the output of the
corresponding NOR gate and its negation, respectively);
\item
inputs of $C$ correspond to tuples of \emph{input}: attribute \emph{Val}
in a tuple of \emph{input} represents the truth assignment to the input
$x_{I\!D\!I\!n\!p\!u\!t}$;
\item
each tuple in \emph{gateInput} represents an input of the gate
identified by \emph{IDGate}.
In particular, \emph{IDIngoing} refers to either a gate identifier or
an input identifier; attribute \emph{Val} is a copy of the truth value of the
specified ingoing gate or input.
\end{enumerate}

We consider the database instance $D$ where the relations defined above are
populated as follows.
For each input $x_i$ in $C$ we insert the tuple $input(id(x_i),-1)$
into $D$, and for each gate $g$ in $C$ we insert the tuple
$gate(id(g),-1,-1)$, where function $id(x)$ assigns a unique identifier
to its argument (we assume that gate identifiers are distinct from input
identifiers, and that the output gate of $C$ is assigned the identifier
$0$).
Moreover, for each edge in $C$ going from $g'$ to the gate $g$ (where
$g'$ is either a gate or an input of $C$), the tuple
$gateInput(id(g),id(g'), -1)$ is inserted into $D$.
Assume that $\mathcal{M}_{gate}=\{norVal, orVal\}$, $\mathcal{M}_{gateInput}=\{Val\}$,
$\mathcal{M}_{input}=\{Val\}$.
In the following, we will define aggregate constraints to force measure
attributes of all tuples to be assigned either $1$ or $0$, representing
the truth value \emph{true} and \emph{false}, respectively.
The initial assignment (where every measure attribute is set to $-1$)
means that the truth values of inputs and gate outputs is undefined.

Consider the following aggregation functions:\\

\noindent
{\small
\begin{tabular}{ll}
$\begin{array}{rl}
N\!O\!RVal(X)=  & \tt SELECT \ Sum(norVal)\\
            & \tt FROM\ gate\\
            & \tt WHERE\ (IDGate\!=\! X)
\end{array}$
&
$\begin{array}{rl}
O\!RVal(X)=   & \tt SELECT\ Sum(orVal)\\
            & \tt FROM\ gate\\
            & \tt WHERE\ (IDGate\!=\!X)
\end{array}$\\ \\
$\begin{array}{rl}
IngoingVal(X,Y)=  & \tt SELECT \ Sum(Val)\\
                & \tt FROM\ gateInput\\
                & \tt WHERE\ (IDGate\!=\! X)\\
                & \tt \ \ \ \ \ AND\ (IDIngoing\!=\! Y)
\end{array}$
&
$\begin{array}{rl}
IngoingSum(X)=  & \tt SELECT \ Sum(Val)\\
                & \tt FROM\ gateInput\\
                & \tt WHERE\ (IDGate\!=\! X)
\end{array}$
\\ \\

$\begin{array}{rl}
InputVal(X)= & \tt SELECT \ Sum(Val)\\
             & \tt FROM\ Input\\
             & \tt WHERE\ (IDInput\!=\!X)
\end{array}$
&
$\begin{array}{rl}
ValidInput(\ )= & \tt SELECT \ Sum(1)\\
                    & \tt FROM\ input\\
                    & \tt WHERE\ (Val\!\neq 0) \\
                    & \tt \ \ \ \ \ AND \  (Val\!\neq 1)\\
\end{array}$\\ \\
\multicolumn{2}{l}{
$\begin{array}{rl}
ValidGate(\ )= & \tt SELECT \ Sum(1)\\
              & \tt FROM\ gate\\
              & \tt WHERE\ (orVal\!\neq 0 \ AND \ orVal\!\neq 1)\\
              & \tt \ \ \ OR\ (norVal\!\neq 0 \ AND \ norVal\!\neq 1)\\
\end{array}$}
\\ \\
\end{tabular}
}

Therein:
$N\!O\!RVal(X)$ and $O\!RVal(X)$ return the truth value of the gate $X$
and its opposite, respectively;
$IngoingVal(X,Y)$ returns, for the gate with identifier $X$, the truth
value of the ingoing gate or input having identifier $Y$;
$IngoingSum(X)$ returns the sum of the truth values of the inputs of the
gate $X$;
$InputVal(X)$ returns the truth assignment of the input $X$;
$ValidInput(\ )$ returns $0$ iff there is no tuple in relation $input$
where attribute $Val$ is neither $0$ nor $1$, otherwise it returns a number
greater than $0$;
likewise, $ValidGate(\ )$ returns $0$ iff there is no tuple in relation $gate$
where attributes $norVal$ or $orVal$ are neither $0$ nor $1$ (otherwise it
returns a number greater than $0$).

Consider the following aggregate constraints on $\mathcal{D}$:\vspace*{-2mm}
\begin{enumerate}
\item
$ValidInput(\ ) + ValidGate(\ ) =0$, which entails that only $0$ and $1$ can
be assigned either to attributes $orVal$ and $norVal$ in relation $gate$, and
to attribute $Val$ in relation $input$;
\item
$gate(X,\_,\_) \Rightarrow O\!RVal(X)+N\!O\!RVal(X)=1$, which says that for
each tuple representing a NOR gate, the value of $orVal$ must be complementary
to $norVal$;
\item
$gate(X,\_,\_) \Rightarrow O\!RVal(X)-IngoingSum(X) \leq 0$, which says that
for each tuple representing a NOR gate, the value of $orVal$ cannot be greater
than the sum of truth assignments of its inputs (i.e. if all inputs are $0$,
$orVal$ must be $0$ too);
\item
$gateInput(X,Y,\_) \Rightarrow IngoingVal(X,Y)-O\!RVal(X) \leq 0$, which implies
that, for each gate $g$, attribute $orVal$ must be $1$ if at least one input of $g$
has value $1$;
\item
$gateInput(X,Y,\_) \Rightarrow IngoingVal(X,Y)-N\!O\!RVal(Y)-InputVal(Y)=0$,
which imposes that the attribute $Val$ in each tuple of $gateInput$ is the
same as the truth value of either the ingoing gate or the ingoing input.
\end{enumerate}\vspace*{-2mm}

Observe that $D$ does not satisfy these constraints, but every repair of $D$
corresponds to a valid truth assignment of $C$.

Let $\mathcal{AC}$ be the set of aggregate constraints consisting of
constraints $1$-$5$ defined above plus constraint $N\!O\!RVal(0)=1$ (which
imposes that the truth value of the output gate must be \emph{true}).
Therefore, deciding whether there is a truth assignment which
evaluates $C$ to $true$ is equivalent to asking whether if there is a
repair $\rho$ for $D$ w.r.t. $\mathcal{AC}$.
\qed
\end{proof}

\noindent
\textbf{Remark.}
Theorem \ref{theo:repexistence} states that the repair existence problem
is decidable.
This result, together with the practical usefulness of the considered class
of constraints, makes the complexity analysis of finding consistent answers
on inconsistent data interesting.
Basically decidability results from the linear nature of the considered
constraints.
If products between two attributes were allowed as attribute expressions,
the repair-existence problem would be undecidable (this can be proved
straightforwardly, since this form of non-linear constraints is more
expressive than those introduced in \cite{Bert*05}, where the corresponding
repair-existence problem was shown to be undecidable).
However, observe that occurrences of products of the form $A_i\times A_j$
in attribute expressions can lead to undecidability only if both $A_i$ and
$A_j$ are measure attribute.
Otherwise, this case is equivalent to products of the form $c\times A$, which
can be expressed in our form of aggregate constraints.

\subsection{Minimal repairs}
Theorem \ref{theo:repexistence} deals with the problem of deciding whether
a database $D$ violating a set of aggregate constraints $\mathcal{AC}$
can be repaired.
If this is the case, different repairs can be performed on $D$ yielding a new
database consistent w.r.t. $\mathcal{AC}$, although not all of them can be
considered ``reasonable".
For instance, if a repair exists for $D$ changing only one value in one tuple
of $D$, any repair updating all values in all tuples of $D$ can be reasonably
disregarded.
To evaluate whether a repair should be considered ``relevant" or not,
we introduce two different ordering criteria on repairs, corresponding to the
comparison operators `$\leq_{\mbox{\scriptsize \emph{set}}}$' and `$\leq_{\mbox{\scriptsize \emph{card}}}$'.
The former compares two repairs by evaluating whether one of the two performs
a subset of the updates of the other.
That is, given two repairs $\rho_1$, $\rho_2$, we say that $\rho_1$ precedes
$\rho_2$ ($\rho_1\leq_{\mbox{\scriptsize \emph{set}}} \rho_2$) iff $\lambda(\rho_1)\subseteq \lambda(\rho_2)$.
The latter ordering criterion states that a repair $\rho_1$ is preferred
w.r.t. a repair
$\rho_2$ ($\rho_1 \leq_{\mbox{\scriptsize \emph{card}}} \rho_2$) iff $|\lambda(\rho_1)|\leq |\lambda(\rho_2)|$,
that is if the number of changes issued by $\rho_1$ is less than $\rho_2$.

%\disc{QUI OCCORRE CHIARIRE LE DUE SEMANTICHE: SE NE POTREBBERO DEFINIRE ALTRE
%(PER ESEMPIO: CONSIDERARE I CAMBI ANZICH\`{E} LE CIFRE CAMBIATE) MA QUESTE
%CI SONO SEMBRATE LE PIU' INTUITIVE.}
%\disc{INOLTRE BISOGNA RICOLLEGARCI AL RELATED WORK SU RIPARAZIONI NEL CAMPO
%RELAZIONALE: SPIEGARE CHE LI' SOLITAMENTE SI USA  UN CONCETTO DIVERSO DI MINIMALITA'}

Observe that $\rho_1\!\!<_{\mbox{\scriptsize \emph{set}}}\!\rho_2$ implies
$\rho_1\!\!<_{\mbox{\scriptsize \emph{card}}}\!\rho_2$, but the vice versa does
not hold, as it can be the case that repair $\rho_1$
changes a set of values $\lambda(\rho_1)$ which is not subset of $\lambda(\rho_2)$,
but having cardinality less than $\lambda(\rho_2)$.

\begin{example}
\label{ex:secondoesempiorepair}
Another repair for \emph{CashBudget}  is
$\rho' = \{ \langle t_1, \mbox{\emph{\small Value}}, 130\rangle, \langle t_2, \mbox{\emph{\small Value}}, 70\rangle,$
$ \langle t_3, \mbox{\emph{\small Value}}, 190\rangle \}$,
where $t_1=\mbox{CashBudget}(\mbox{ 2003, Receipts, cash sales, det, 100})$,
$t_2=\mbox{CashBudget}(\mbox{ 2003, Disbursements, long-term financing, det, 40})$,
and
$t_3=\mbox{CashBudget}$ $(\mbox{ 2003, Disbursements, total disbursements, aggr, 160})$.\\
Observe that $\rho<_{\mbox{\scriptsize card}}\rho'$, but not $\rho<_{set}\rho'$
(where $\rho$ is the repair defined in Example \ref{ex:primoesempiorepair}).
\end{example}

\begin{definition}[Minimal repairs]
\label{def:setminimalrepair}
Let $\mathcal{D}$ be a database scheme, $\mathcal{AC}$ a set of aggregate
constraints on $\mathcal{D}$, and $D$ an instance of $\mathcal{D}$.
A repair $\rho$ for $D$ w.r.t.  $\mathcal{AC}$ is a \emph{set}-minimal
repair [resp. \emph{card}-minimal repair] iff there is no repair $\rho'$
for $D$ w.r.t. $\mathcal{AC}$ such that $\rho' <_{\mbox{\scriptsize set}} \rho$
[resp. $\rho' <_{\mbox{\scriptsize card}} \rho$].
%\begin{list}{-}{\leftmargin .1cm \itemsep 0cm \parsep 0cm}
%\item
%A repair $\rho$ for $D$ w.r.t.  $\mathcal{AC}$ is a \emph{set}-minimal
%repair iff there is no repair $\rho'$ for $D$ w.r.t. $\mathcal{AC}$ such
%that $\rho' <_{\mbox{\scriptsize set}} \rho$.
%\item
%A repair $\rho$ for $D$ w.r.t.  $\mathcal{AC}$ is a \emph{card}-minimal
%repair iff there is no repair $\rho'$ for $D$ w.r.t.  $\mathcal{AC}$
%such that $\rho' <_{\mbox{\scriptsize card}} \rho$.
%\end{list}
\end{definition}

\begin{example}
Repair $\rho$ of Example \ref{ex:primoesempiorepair} is minimal under both
the \emph{set}-minimal and the \emph{card}-minimal semantics, whereas $\rho'$ defined in
Example \ref{ex:secondoesempiorepair} is minimal only under the \emph{set}-minimal
semantics.\\
Consider the repair $\rho''$ consisting of the following updates:
$\rho''=\{
\langle t_1, \mbox{\emph{Value}}, 110\rangle,$ $
\langle t_2, \mbox{\emph{Value}}, 110\rangle,
\langle t_3, \mbox{\emph{Value}}, 220\rangle
\}$
where:
$t_1=\mbox{CashBudget}(\mbox{ 2003, Receipts, cash sales,}$ $\mbox{det, 100})$,
$t_2=\mbox{CashBudget}(\mbox{ 2003, Receipts, receivables, det, 120})$,
$t_3=\mbox{CashBudget}($ $\mbox{ 2003, Receipts, total cash receipts, aggr, 250})$.\\
The strategy adopted by $\rho''$ can be reasonably disregarded, since
the only atomic update on tuple $t_3$ suffices to make $D$ consistent.
In fact, $\rho''$ is not minimal neither under the \emph{set}-minimal semantics
( as $\lambda(\rho)\subset\lambda(\rho'')$ and thus $\rho\!\!<_{\mbox{\scriptsize set}}\!\rho''$)
nor under the \emph{card}-minimal one.
\end{example}

Given a database $D$ which is not consistent w.r.t. a set of aggregate constraints
$\mathcal{AC}$, different \emph{set}-minimal repairs (resp. \emph{card}-minimal repairs)
can exist on $D$.
In our running example, repair $\rho$ of Example \ref{ex:primoesempiorepair} is the
unique \emph{card}-minimal repair, and  both $\rho$ and $\rho'$ are \emph{set}-minimal
repairs (where $\rho'$ is the repair defined in Example \ref{ex:secondoesempiorepair}).
The set of \emph{set}-minimal repairs and the set of \emph{card}-minimal repairs
will be denoted, respectively, as $\rho^{\mbox{\scriptsize \emph{set}}}_M$ and
$\rho^{\mbox{\scriptsize \emph{card}}}_M$.
%VERIFICHIAMO SE VIENE USATA QUESTA NOTAZIONE

\begin{theorem}[Minimal-repair checking]
\label{theo:minimalrepairchecking}
Let $\mathcal{D}$ be a database scheme, $\mathcal{AC}$ a set of aggregate
constraints on $\mathcal{D}$, and $D$ be an instance of $\mathcal{D}$ such that
$D \not\models \mathcal{AC}$.
Given a repair $\rho$ for $D$ w.r.t. $\mathcal{AC}$, deciding whether
$\rho$ is minimal (under both the \emph{card}-minimality and \emph{set}-minimality
semantics) is coNP-complete.
\end{theorem}
%\bproof
\begin{proof}
(Membership)
A polynomial size witness for the complement of the problem of deciding
whether $\rho \in \rho^{\mbox{\scriptsize \emph{set}}}_M$
[resp. $\rho \in \rho^{\mbox{\scriptsize \emph{card}}}_M$] is a repair
$\rho'$ such that $\rho'<_{\mbox{\scriptsize \emph{set}}} \rho$
[resp. $\rho'<_{\mbox{\scriptsize \emph{card}}} \rho$].
From Lemma \ref{lem:dimpolrepair} we have that $\rho'$ can be found among
repairs having polynomial size w.r.t. $D$.\\
(Hardness)
We show a reduction of \textsc{minimal model checking}~(\textsc{mmc})~\cite{Cad*96}
to our problem.
Consider an instance $\langle f, M \rangle$ of \textsc{mmc}, where
$f$ is a propositional formula and $M$ a model for $f$.
Formula $f$ can be translated into an equivalent boolean circuit $C$ using
only NOR gates, and $C$ can be represented as shown in the hardness proof
of Theorem~\ref{theo:repexistence}.
Therefore, we consider the same database scheme $\mathcal{D}$ and the same
set of aggregate constraints $\mathcal{AC}$ on $\mathcal{D}$ as those in the
proof of Theorem~\ref{theo:repexistence}.
Let $D$ be the instance of $\mathcal{D}$ constructed as follows.
For each input $x_i$ in $C$ we insert the tuple $input(id(x_i),0)$ into $D$.
Then, as for the construction in the hardness proof of
Theorem~\ref{theo:repexistence}, for each gate $g$ in $C$ we insert the
tuple $gate(id(g),-1,-1)$ into $D$, and for each edge in $C$ going from
$g'$ to the gate $g$ (where $g'$ is either a gate or an input of $C$), the
tuple $gateInput(id(g),id(g'), -1)$ is inserted into $D$.

Observe that any repair for $D$ must update all measure attributes in $D$
with value $-1$.
Therefore, given two repairs $\rho'$, $\rho''$, it holds that
for each $<t, A>\, \in \, (\lambda(\rho')\, \triangle\, \lambda(\rho''))$, $t$
is a tuple of $input$ and $A =Val$.

Obviously, a repair $\rho$ for $D$ exists, consisting of the following updates:
1) attribute $Val$ is assigned $1$ in every tuple of $input$ corresponding
to an atom in $f$ which is true in $m$;
2) attributes $norVal$, $orVal$ in $gate$ and $Val$ in $gateInput$ are
updated accordingly to updates described above.
Basically, such a constructed repair $\rho$ corresponds to $M$ (we say that
a repair corresponds to a model if it assigns $1$ to attribute $Val$ in the
tuples of $input$ corresponding to the atoms which are true in the model,
$0$ otherwise).

If $M$ is not a minimal model for $f$, then there exists a model $M'$ such
that $M' \subset M$ (i.e. atoms which are true in $M'$ are a proper subset of atoms
which are true in $M$).
Then, the repair $\rho'$ corresponding to $M'$ satisfies $\rho' <_{set} \rho$.
Vice versa, if there exists a repair $\rho'$ such that $\rho' <_{set} \rho$,
then the model $M'$ corresponding to $\rho'$ is a proper subset of $M$, thus $M$ is
not minimal.
This proves that $M$ is a minimal model for $f$ iff $\rho$ is a minimal repair
(under \emph{set}-minimal semantics) for $D$ w.r.t $\mathcal{AC}$.

Proving hardness under \emph{card}-minimal semantics can be accomplished as follows.
First, a formula $f_M$ is constructed from $f$ by replacing, for each atom
$a \not\in M$, each occurrence of $a$ in $f$ with the contradiction
$( a \wedge \neg a )$.
Then, an instance $D$ of $\mathcal{D}$ is constructed corresponding to formula
$f_M$ with the same value assignments as before (attribute $Val$ in all the
tuples of $input$ are set to $0$, and all the other measure attributes are
set to $-1$).

$M$ is a model for both $f$ and $f_M$, and it is minimal for $f$ iff it is
minimum for $f_M$.
In fact, if $M$ is minimal for $f$ there is no subset $M'$ of $M$ which is
a model of $f$.
Then, assume that a model $M''$ for $f_M$ exists, such that $|M''|<|M|$.
Then, also $M'''=M''\cap M$ is a model for $f_M$, implying that $M'''$ is
a model for $f$, which is a contradiction (as $M''' \subset M$).
On the other hand, if $M$ is minimum for $f_M$ then $M$ must be minimal for $f$.
Otherwise, there would exist a model $M'$ for $f$ s.t. $M' \subset M$.
However $M'$ is also a model for $f_M$, which is a contradiction, as $|M'|<|M|$.

Let $\rho$ be the repair of $D$ w.r.t. $\mathcal{AC}$ corresponding to $M$.
If $M$ is not minimum, then there exists $M'$ (with $|M'| < |M|$) which is a model
for $f_M$.
Therefore the repair $\rho'$ corresponding to $M'$ satisfies $\rho' <_{card} \rho$.
Vice versa, if a repair $\rho'$ for $D$ w.r.t. $\mathcal{AC}$ exists such that
$\rho' <_{card} \rho$, then the model $M'$ corresponding to $\rho'$ is such that
$|M'| < |M|$, thus $M$ is not minimum for $f_M$.
This proves that $M$ is a minimal model for $f$ iff there is no repair $\rho'$ for
$D$ w.r.t. $\mathcal{AC}$ such that $\rho' <_{card} \rho$.
\qed
\end{proof}

\noindent\textbf{\emph{Set-minimality vs card-minimality}}\ \\
Basically, both the \emph{set}-minimal and the \emph{card}-minimal semantics aim at considering
``reasonable" repairs which preserve  the content of the input database as much
as possible.
To the best of our knowledge the notion of repair minimality based on the number
of performed updates has not been used in the context of relational data violating
``non-numerical" constraints (such as keys, foreign keys, and functional
dependencies).
In this context, most of the proposed approaches consider repairs consisting of
deletions and insertions of tuples, and preferred repairs are those consisting
of minimal sets of insert/delete operations.
In fact, the \emph{set}-minimal semantics is more natural than the \emph{card}-minimal one
when no hypothesis can be reasonably formulated to ``guess" how data
inconsistency occurred, which is the case of previous works on database-repairing.
As it will be clear in the following, in the general case, the adoption of the
\emph{card}-minimal semantics could make reasonable sets of delete/insert operations to
be not considered as candidate repairs, even if they correspond to error configurations
which cannot be excluded.

For instance, consider a relational scheme \emph{Department(Name, Area, Employers, Category)}
where the following functional dependencies are defined:
$FD_1: Area \rightarrow Employers$ (i.e. departments having the same area must have the
same number of employers)
and
$FD_2: Employers \rightarrow Category$ (i.e. departments with the same number of
employers must be of the same category).
Consider the following relation:

\begin{center}
{
\small
\hspace*{6mm}
\begin{tabular}{||l|c|c|c||r}
  \hhline{|t:====:t|}
  \raisebox{0cm}[3mm][1mm]{\textbf{\textit{Department}}} & \textbf{\textit{Area}} & \textbf{\textit{Employers}} & \textbf{\textit{Category}} & \\
  \hhline{|:====:|}
  \raisebox{0cm}[2.9mm][1mm]{$D_1$} & 100       & 24 & A &  $\longrightarrow t_1$\\
  \hhline{||----||}
  \raisebox{0cm}[2.9mm][1mm]{$D_2$} & 100       & 30     & B & $\longrightarrow t_2$\\
  \hhline{||----||}
  \raisebox{0cm}[2.9mm][1mm]{$D_3$} & 100       & 30    & B & $\longrightarrow t_3$\\

\hhline{|b:====:b|}
\end{tabular}
}
\end{center}

Relation above does not satisfy $FD_1$, as the three departments occupy
the same area but do not have the same number of employers.
Suppose we are using a repairing strategy based on deletions and insertions
of tuples.
Different repairs can be adopted.
For instance, if we suppose that the inconsistency arises as tuple $t_1$
contains wrong information, \emph{Department} can be repaired by only
deleting $t_1$.
Otherwise, if we assume that $t_1$ is correct, a possible repair
consists of deleting $t_2$ and $t_3$.
If the \emph{card}-minimal semantics is adopted, the latter strategy will be
disregarded, as it performs two deletions, whereas the former deletes
only one tuple.
On the contrary, if the \emph{set}-minimal semantics is adopted, both the two strategies
define minimal repairs (as the sets of tuples deleted by each of these strategies
are not subsets of one another).
In fact, if we do not know how the error occurred, there is no reason to assume
that the error configuration corresponding to the second repairing strategy is
not possible.
Indeed, inconsistency could be due to integrating data coming from different
sources, where some sources are not up-to-date. However, there is no good reason to
assume that the source which contains the smallest number of tuples
is the one that is up to date. See \cite{Len02} for a survey on inconsistency
due to data integration.
%For instance, it can be the case that tuple $t_1$ was extracted from a source
%where all tuples whose attribute \emph{Area} is 100 have attribute \emph{Employers}
%set to 24, whereas $t_2$ and $t_3$ were extracted from a source where all tuples
%whose attribute \emph{Area} is 100 have attribute \emph{Employers} set to 30,
%as the two sources describe two organizations of departments corresponding to
%different time periods.

Likewise, the \emph{card}-minimal semantics could disregard reasonable
repairs also in the case that a repairing strategy based on
updating values instead of deleting/inserting whole tuples is
adopted$\!$
\footnote{Value updates cannot be necessarily simulated
as a sequence deletion/insertion, as this might not be minimal under
set inclusion.}.
For instance, if we suppose that the inconsistency arises as the
value of attribute \emph{Area} is wrong for either $t_1$ or both
$t_2$ and $t_3$, \emph{Department} can be repaired by replacing
the \emph{Area} value for either $t_1$ or both $t_2$ and $t_3$
with a value different from $100$.
Otherwise, if we assume that the \emph{Area} values for all the tuples
are correct, \emph{Department} can be repaired w.r.t.
$FD_1$ by making the \emph{Employers} value of $t_1$ equal
to that of $t_2$ and $t_3$.
Indeed this update yields a relation which does not satisfy
$FD_2$ (as $t_1[\mbox{\emph{Category}}]\neq t_2[\mbox{\emph{Category}}]$)
so that another value update is necessary in order to make it
consistent.
Under the \emph{card}-minimal semantics the latter strategy is disregarded,
as it performs more than one value update, whereas the former changes
only the \emph{Area} value of one tuple.
On the contrary, under the \emph{set}-minimal semantics both the two strategies
define minimal repairs (as the sets of updates issued by each of these
strategies are not subsets of one another).
As for the case explained above, disregarding the second repairing strategy
is arbitrary, if we do not know how the error occurred.

Our framework addresses scenarios where also \emph{card}-minimal semantics can be
reasonable.
For instance, if we assume that integrity violations are generated while acquiring
data by means of an automatic or semi-automatic system (e.g. an OCR
digitizing a paper document, a sensor monitoring atmospheric conditions, etc.),
focusing on error configurations which can be repaired with the minimum number of
updates is well founded.
Indeed this corresponds to the case that the acquiring system made the minimum
number of errors (e.g. bad symbol-recognition for an OCR, sensor troubles, etc.),
which can be considered the most probable event.

In this work we discuss the existence of repairs, and their computation
under both \emph{card}-minimal and \emph{set}-minimal semantics.
The latter has to be preferred when no warranty is given on the accuracy of
acquiring tools, and, more generally, when no hypothesis can be formulated on
the cause of errors.

% IN SOSTANZA: PROPONIAMO LA CARD MINIMAL NON PERCHè E' PIU' ADATTA A GESTIRE I REPAIR
% CON SOSTITUZIONI, MA PERCHE' E' PIU' ADATTA ALLO SCENARIO

\subsection{Consistent query answers}
In this section we address the problem of extracting reliable information
from data violating a given set of aggregate constraints.
%As shown in the previous section, several repairs can be issued on inconsistent
%data to make them satisfying the associated constraints.
We consider boolean queries checking whether a given tuple belongs to
a database, and adopt the widely-used notion of consistent query answer
introduced in \cite{Are*99}.

\begin{definition}[Query]
\label{def:query}
A \emph{query} over a database scheme $\mathcal{D}$ is a ground atom
of the form $R(v_1, \dots, v_n)$, where $R(A_1, \dots, A_n)$ is
a relational scheme in $\mathcal{D}$.
\end{definition}

\begin{definition}[Consistent query answer]
\label{def:certainanswer}
Let $\mathcal{D}$ be a database scheme,
$D$ be an instance of $\mathcal{D}$,
$\mathcal{AC}$ be a set of aggregate constraints on $\mathcal{D}$ and
$q$ be a query over $\mathcal{D}$.
The \emph{consistent query answer} of $q$ on $D$ under the \emph{set}-minimal semantics
[resp. \emph{card}-minimal semantics] is true iff $q \in \rho(D)$ for
each $\rho \in \rho_M^{\mbox{\scriptsize  set}}$
[resp. for each $\rho \in \rho_M^{\mbox{\scriptsize card}}$].
%\begin{itemize}
%\item
%The certain answer to $q$ on $D$ under the \emph{set}-minimal semantics
%(denoted as $q^{set}(D)$)
%is true iff $q(R(D))$ is true for each \emph{set}-minimal repair of $D$.
%\item
%The certain answer to $q$ on $D$ under the \emph{card}-minimal semantics
%(denoted as $q^{card}(D)$)
%is true iff $q(R(D))$ is true for each \emph{card}-minimal repair of $D$.
%\end{itemize}
\end{definition}

The consistent query answers of a query $q$ issued on the database $D$
under the \emph{set}-minimal and \emph{card}-minimal semantics will be
denoted as $q^{set}(D)$ and $q^{card}(D)$, respectively.

%\begin{theorem}\label{theo:consans}
%Let $\mathcal{D}$ be a database scheme, $D$ be an instance of $\mathcal{D}$,
%$\mathcal{AC}$ be a set of aggregate constraints on $\mathcal{D}$ and
%$q$ be a query over $D$.
%Deciding whether $q^{card}(D) = true$ is in $\Delta_2^p$,
%and deciding whether $q^{set}(D) = true$ is in $\Pi_2^p$.
%Both these problems are $DP$-hard.
%\end{theorem}
%%\bproof
%\begin{proof}
%(sketch)
%Membership in $\Pi_2^p$ can be proved by reasoning as for Theorem
%\ref{theo:repexistence}, by exploiting a result similar to that of
%Lemma \ref{lem:dimpolrepair} (it can be proved that if there exists
%a repair $\rho$ s.t. $q(\rho(D))$ is true, then there exists a repair
%$\rho'$ having polynomial size w.r.t. $q$ and $D$ s.t.
%$\lambda(\rho')\subseteq \lambda(\rho)$ ).\\
%Under the \emph{card}-minimal semantics, membership in $\Delta_2^p$
%derives from the fact that repairs on $D$ can be partitioned into the
%two sets $T$ and $F$ consisting of all repairs $\rho_i$ s.t.
%$q(\rho_i(D)) = \mbox{true}$ and, respectively,
%$q(\rho_i(D)) = \mbox{false}$.
%Let $MinSize(T)= min_{\rho \in T} (|\lambda(\rho)|)$, and
%$MinSize(F)= min_{\rho \in F} (|\lambda(\rho)|)$.
%It can be shown that $q^{card}(D) = \mbox{true}$ iff
%$MinSize(T)<MinSize(F)$.
%Both $MinSize(T)$ and $MinSize(F)$ can be evaluated by a logarithmic number of NP-oracle invocations.\\
%We have proved DP-hardness by reducing \textsc{Sat-Unsat} to our
%problem.
%The reduction is rather elaborated and thus omitted for the sake
%of brevity.
%\qed
%\end{proof}
%%\eproof

\begin{theorem}[Consistent query answer under \emph{set}-minimal semantics]
\label{theo:consAnsSet}
Let $\mathcal{D}$ be a database scheme, $D$ be an instance of $\mathcal{D}$,
$\mathcal{AC}$ be a set of aggregate constraints on $\mathcal{D}$ and
$q$ be a query over $D$.
Deciding whether $q^{set}(D) = true$ is $\Pi_2^p$-complete.
\end{theorem}

\begin{proof}
See appendix.
\qed
\end{proof}

\begin{theorem}[Consistent query answer under \emph{card}-minimal semantics]\label{theo:consAnsCard}
Let $\mathcal{D}$ be a database scheme, $D$ be an instance of $\mathcal{D}$,
$\mathcal{AC}$ be a set of aggregate constraints on $\mathcal{D}$ and
$q$ be a query over $D$.
Deciding whether $q^{card}(D) = true$ is $\Delta_2^p[log \ n]$-complete.
\end{theorem}

\begin{proof}
See appendix.
\qed
\end{proof}

\section*{Conclusions and Future Work}
We have addressed the problem of repairing and extracting reliable
information from numerical databases violating aggregate constraints,
thus filling a gap in previous works dealing with inconsistent data,
where only traditional forms of constraints (such as keys, foreign keys,
etc.) were considered.
In fact, aggregate constraints frequently occur in many real-life
scenarios where guaranteeing the consistency of numerical data is
mandatory.
In particular, we have considered aggregate constraints defined as
sets of linear inequalities on aggregate-sum queries on input data.
For this class of constraints we have characterized the complexity of
several issues related to the computation of consistent query answers.

Future work will be devoted to the identification of decidable cases
when more expressive forms of constraint are adopted, that allow products
between attribute values (as explained in the paper, enabling non-linear
forms of aggregate expressions makes the repair-existence problem undecidable
in the general case).
Moreover the design of efficient algorithms for computing consistent answers
will be addressed.

\renewcommand{\baselinestretch}{.96}

\newpage
\section*{Appendix: Proofs of theorems}

\textbf{Theorem \ref{theo:consAnsSet}. }
\emph{Let $\mathcal{D}$ be a database scheme, $D$ be an instance of $\mathcal{D}$,
$\mathcal{AC}$ be a set of aggregate constraints on $\mathcal{D}$ and
$q$ be a query over $D$.
Deciding whether $q^{set}(D) = true$ is $\Pi_2^p$-complete.
}

\begin{proof}
\noindent (Membership)
Membership in $\Pi_2^p$ can be proved by reasoning as for Theorem~
\ref{theo:repexistence}, by exploiting a result similar to that of
Lemma \ref{lem:dimpolrepair} (it can be proved that if there is
a repair $\rho$ s.t. $q(\rho(D))$ is true, then there is a repair
$\rho'$ having polynomial size w.r.t. $q$ and $D$ s.t.
$\lambda(\rho')\subseteq \lambda(\rho)$ ).

\noindent (Hardness)
Hardness can be proved by showing a reduction from the following
implication problem in the context of propositional logic over a
finite domain $V$, which was shown to be $\Pi_2^p$-complete in
\cite{Gott*92}:
``\emph{given an atomic knowledge base $T=\{a_1, \dots, a_n\}$,
where $a_1, \dots, a_n$ are atoms of $V$, an atom $Q\in T$ and
a formula $p$ on $V$, decide whether $Q$ is derivable from every
model in $T \circ_S p$}",
where $T \circ_S p$ is the updated (or revised) knowledge base
according to the Satoh's revision operator.

Informally, Satoh's revision operator $\circ_S$ selects the models of
$p$ that are ``closest" to models of $T$: closest models are those
whose symmetric difference with models of $T$ is minimal under
set-inclusion semantics.
In order to define formally the semantics of $\circ_S$ we first
introduce some preliminaries.
Let $Mod(p)$ be the set of models of a formula $p$.
%Given two models $M$, $M'$, $M \triangle M'$  denotes their symmetric
%difference $( M \cup M' ) \setminus ( M \cap M' )$.
Let
$\triangle^{min}(T,p)= min_\subseteq ( \{ M \triangle M' : M \in Mod(p) ,\  M' \in Mod(T) \} )$,
that is the family of $\subseteq$-minimal sets obtained as symmetric
difference between models of $p$ and $T$.
The semantics of Satoh's operator (i.e. the set of models of the knowledge
base $T$ revised according to the formula $p$) is  defined as follows:\\
$Mod(T \circ_S p)=\{~M~\in~Mod(p)~: \exists M' \in Mod(T) \ s.t. \  M \triangle M' \in \triangle^{min}(T,p) \}$.

In the following the set of atoms occurring in $p$ will be denoted as
$V(p)$.
$\Pi_2^p$-completeness of the implication problem was shown to hold also
if $V(p)~\subseteq~T$~\cite{Gott*92}:
we consider this case in our proof.
Observe that the definition of $\circ_S$ entails that for each
$M \in \triangle^{min}(T,p)$ it holds that $M \subseteq T \cap V(p)$,
thus $M$ is a subset of $T$.
%The implication problem we consider exactly corresponds to evaluating a
%counterfactual according the Satoh's revision semantic.
%Counterfactual is a conditional statement of the form
%``if $p$ were true, then $q$ would hold'', where $p$ is know or assumed to be false.
%According to the \emph{Ramsey Test} \cite{Gar*88}, evaluating such a counterfactual
%(denoted as $p >_S q$) in a given knowledge base $T$ is equivalent to test whether $q$
%is logical consequence of $T \circ_S p$.

We now consider an instance $<T, p, Q>$ of implication problem, where
$T$ is the atomic knowledge base $\{ a_1, \dots , a_n \}$,
$p$ is a propositional formula (with $V(p) \subseteq T$), and $Q$ is
an atom in $T$.

Let $C_p$ be a boolean circuit equivalent to $p$.
We consider the database scheme $\mathcal{D}$ introduced in the
hardness proof of Theorem \ref{theo:repexistence}.
Moreover, we consider an instance $D$ which is the translation of
$C_p$ obtained in the same way as Theorem~\ref{theo:repexistence},
except that:
\begin{itemize}
\item
relation $input$ must contain not only the tuples corresponding to
the inputs of $C_p$ (i.e. the atoms in $V(p)$), but also the tuples
corresponding to the atoms of $T\setminus V(p)$;
\item
for each tuple inserted in relation $input$, attribute $Val$ is set to
$1$, which means assigning \emph{true} to all the atoms of $T$.
\end{itemize}
Recall that measure attributes in the tuples of relations $gate$ and
$gateInput$ are set to $-1$ (corresponding to an undefined truth value).

Let $\mathcal{AC}$ be the same set of constraints used in the proof
of Theorem~\ref{theo:repexistence}.
As explained in the hardness proof of Theorem~\ref{theo:repexistence},
$\mathcal{AC}$ defines the semantics of $C_p$ and requires that $C_p$
is true.
Note that every repair $\rho$ for $D$ w.r.t. $\mathcal{AC}$ must update
all measure attributes that initially are set to $-1$ in $D$.
Therefore, given two repairs $\rho$ and  $\rho'$, they differ only on the
set of atomic updates performed on relation $input$.

Obviously, every \emph{set}-minimal repair of $\rho$ for $D$ w.r.t.
$\mathcal{AC}$ corresponds to a model $M$ in $Mod(T \circ_S p)$, and
vice versa.
In fact, given a \emph{set}-minimal repair $\rho$ for $D$ w.r.t.
$\mathcal{AC}$, a model $M$ for $T \circ_S p$ can be obtained from the
repaired database considering only the tuples in relation $input$
where attribute $Val$ is equal to $1$ after applying $\rho$.
Observe that the set of atoms $M$ corresponding to $\rho$ is a model
$T \circ_S p$, otherwise there would exist $M' \subset M$ with
$M' \in Mod(T \circ_S p)$,
and the repair $\rho'$ corresponding to $M'$ would satisfy $\rho' <_{set} \rho$,
thus contradicting the minimality of $\rho$.
Likewise, it is easy to see that any model in $Mod(T \circ_S p)$ corresponds
to a minimal repair for $D$ w.r.t. $\mathcal{AC}$.

Finally consider the query $q= input(id(Q),1)$.
The above considerations suffice to prove that $Q$ is derivable from every
model in $Mod(T \circ_S p)$ iff $input(id(Q),1)$ is true in $\rho(D)$ for
every \emph{set}-minimal repair $\rho$ for $D$ w.r.t. $\mathcal{AC}$, that
is the consistent answer of $input(id(Q),1)$ on $D$ w.r.t. $\mathcal{AC}$
is true.
\qed
\end{proof}

\noindent
\textbf{Theorem \ref{theo:consAnsCard}. }
\emph{Let $\mathcal{D}$ be a database scheme, $D$ be an instance of $\mathcal{D}$,
$\mathcal{AC}$ be a set of aggregate constraints on $\mathcal{D}$ and
$q$ be a query over $D$.
Deciding whether $q^{card}(D) = true$ is $\Delta_2^p[log \ n]$-complete.
}

\begin{proof}
(\emph{Membership})
Membership in $\Delta_2^p[log \ n]$
derives from the fact that repairs on $D$ can be partitioned into the
two sets $T$ and $F$ consisting of all repairs $\rho_i$ s.t.
$q(\rho_i(D)) = \mbox{\emph{true}}$ and, respectively,
$q(\rho_i(D)) = \mbox{\emph{false}}$.
Let \emph{MinSize}$(T)= min_{\rho \in T} (|\lambda(\rho)|)$, and
$MinSize(F)= min_{\rho \in F} (|\lambda(\rho)|)$.
It can be shown that $q^{card}(D) = \mbox{\emph{true}}$ iff
\emph{MinSize}$(T)<$ \emph{MinSize}$(F)$.
Both \emph{MinSize}$(T)$ and \emph{MinSize}$(F)$ can be evaluated
by a logarithmic number of NP-oracle invocations.\\
(\emph{Hardness}).
Hardness can be proved by showing a reduction from the following implication
problem in the context of propositional logic over a finite domain $V$:
``\emph{given an atomic knowledge base $T$ on $V$, a formula $Q$ on $T$ and
a formula $p$ on $V$, decide whether $Q$ is derivable from every
model in $T \circ_D p$}",
where $T \circ_D p$ is the updated (or revised) knowledge base
according to the Dalal's revision operator.
$\Delta_2^p[log \ n]$-completeness of this problem was shown in
\cite{Gott*92}.

The semantics of Dalal's revision operator is as follows.
The models of $T \circ_D p$ are the models of $p$ whose symmetric
difference with models of $T$ has minimum cardinality w.r.t. all other models of $p$.
%That is, models of $T \circ_D p$ are the models of $p$ having as
%few variables different from any model of $T$ as possible.
More formally, let $|\triangle^{min}(T,p)| = min \{~|M \triangle M'|~: M \in Mod(p) ,\  M' \in Mod(T) \} $,
that is the minimum number of atoms in which models of $T$ and $p$
diverge.
Then models of $T \circ_D p$ are given by:\\
$Mod(T \circ_D p)=\{ M \in Mod(p) : \exists M' \in Mod(T) \ s.t.\
| M \triangle M' | \in | \triangle|^{min}(T,p) \}$.

Consider an instance $<V, T, p, Q>$ of the implication problem, where
$V$ is the finite domain of atoms,
$T$ an atomic knowledge base on $V$,
$p$ a formula on $V$,
and $Q$ a formula on $T$.
Let $V(p)$ and $V(Q)$ denote the set of atoms of $V$ occurring in $p$ and $Q$,
respectively.
Sets $T$, $V(p)$ and $V(Q)$ can be partitioned into $A$, $B$, $C$, $D$, $E$,
as shown in Fig. \ref{fig:teorema4}(a).

\begin{figure}
  \begin{tabular}{c}
  \includegraphics[width=.99\textwidth]{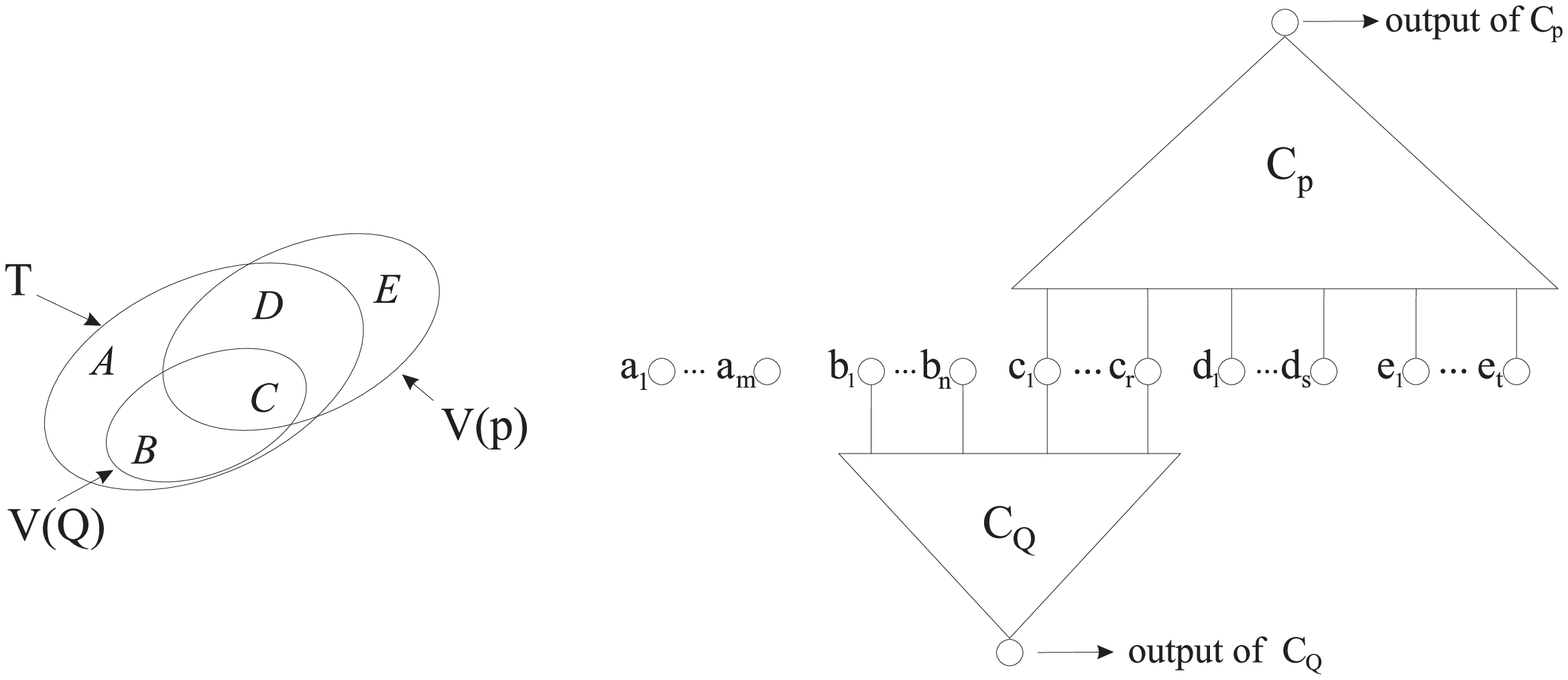}\\
  (a)\hspace*{6cm}(b)
  \end{tabular}
  \caption{(a) The partitioning of $T$, $V(p)$, $V(Q)$; (b) Circuits }
  \label{fig:teorema4}
\end{figure}

Let $C_p$ and $C_Q$ be two boolean circuits equivalent to $p$ and $Q$,
respectively.
$C_p$ and $C_Q$ are reported in Fig. \ref{fig:teorema4}(b), with their
inputs.
In this figure, atoms belonging to $T$, $V(p)$ and $V(Q)$ are represented
as circles, and the two circuits are represented by means of triangles.
In particular, inputs of $C_Q$ are the atoms $b_1, \dots, b_n$ of $B$ and
the atoms $c_1, \dots, c_r$ of $C$, whereas
inputs of $C_Q$ are the atoms $c_1, \dots, c_r$ of $C$,
the atoms $d_1, \dots, d_s$ of $D$, and the atoms $e_1, \dots, e_t$ of $D$.
That is, the atoms of $C$ are inputs of both $C_p$ and $C_Q$.

These circuits can be represented as an instance of the database scheme
$\mathcal{D}$ introduced in the hardness proof of Theorem~\ref{theo:repexistence}.
In particular, we consider an instance $D$ of $\mathcal{D}$
which is the translation of $C_p$ and $C_Q$ obtained in the same way as
Theorem~\ref{theo:repexistence}, except that:
\begin{itemize}
\item
relation $input$ contains a tuple for each atom in
$A \cup B \cup C \cup D \cup E$;
\item
for each tuple inserted in relation $input$, attribute $Val$ is set to
$1$ if it refers to an atom in $T$, $-1$ otherwise.
This means assigning \emph{true} to all the atoms of $T$,
and an undefined truth value to atoms in $E$.
\end{itemize}

Recall that measure attributes in the tuples of relations $gate$ and
$gateInput$ are set to $-1$.

We consider the set of aggregate constraints $\mathcal{AC}$
consisting of constraints 1-5 introduced in the
hardness proof of Theorem~\ref{theo:repexistence}, plus the aggregate
constraint $N\!O\!RVal(id(o_p))=1$, where $id(o_p)$ is the identifier of
the output gate of $C_p$.
%and $id(o_Q)$ is the identifier assigned to the output gate of $C_Q$.
As explained in the hardness proof of Theorem~\ref{theo:repexistence},
$\mathcal{AC}$ defines the semantics of $C_p$ and $C_Q$ and requires
that $C_p$ is true.

Note that every repair $\rho$ for $D$ w.r.t. $\mathcal{AC}$ must update
all value attributes that initially are assigned -1 in $D$.
Therefore, given two repairs $\rho$ and  $\rho'$ for $D$ w.r.t. $\mathcal{AC}$,
they differ only on the number of atomic updates performed on the tuples
of $input$ where $Val$ was set to $1$ in $D$.

Obviously, every \emph{card}-minimal repair of $\rho$ for $D$ w.r.t.
$\mathcal{AC}$ corresponds to a model $M$ in $Mod(T \circ_D p)$, and
vice versa (this can be proven straightforwardly, analogously to the
proof of Theorem~\ref{theo:consAnsSet}, where the correspondence between
\emph{set}-minimal repairs for $D$ and models of $T \circ_S p$ has been shown).

Finally consider the query $q= input(id(o_Q),1)$, where $o_Q$ denotes the
the output gate of $C_Q$.
The above-mentioned considerations suffice to prove that $Q$ is derivable
from every model in $Mod(T \circ_D p)$ iff $input(id(o_Q),1)$ is true in
$\rho(D)$ for every \emph{card}-minimal repair $\rho$ for $D$ w.r.t.
$\mathcal{AC}$, that is the consistent answer of $input(id(o_Q),1)$ on $D$
w.r.t. $\mathcal{AC}$ is true.
\qed
\end{proof}
\end{document}